\documentclass{article}
\usepackage{ijcai26}
\usepackage{times}
\usepackage{soul}
\usepackage{url}
\usepackage[hidelinks]{hyperref}
\usepackage[utf8]{inputenc}
\usepackage[small]{caption}
\usepackage{graphicx}
\usepackage{amsmath}
\usepackage{booktabs}
\usepackage{algorithm}
\usepackage{algorithmic}

\ifdefined
\else

\fi

\title{Landscape-Aware Bandit Hyper-Heuristics for Online Operator Selection in UAV Inspection Routing}

\author{
Junhao Wei$^{1,2}$\and
Yanxiao Li$^1$\and
Yifu Zhao$^1$\and
Qibin He$^1$\and
Haochen Li$^1$\and
Dexing Yao$^1$\and
Baili Lu$^1$\and
Zhenhong Peng$^3$\and
Yapeng Wang$^1$\and
Sio-Kei Im$^4$\And
Xu Yang$^1$\thanks{Corresponding author.}\\
\affiliations
$^1$Faculty of Applied Sciences, Macao Polytechnic University, Macao, 999078, China\\
$^2$Pazhou Lab (Huangpu), Guangzhou, 510555, China\\
$^3$College of Animal Science and Technology, Zhongkai University of Agriculture and Engineering, Guangzhou, 510225, China\\
$^4$Macao Polytechnic University, Macao, 999078, China\\
\emails
xuyang@mpu.edu.mo
}

\begin{document}

\maketitle

\begin{abstract}
UAV multi-site inspection often reduces to choosing a high-quality visiting order after target sites have been extracted from a map.
This paper develops LA-BHH, a landscape-aware bandit hyper-heuristic that learns an operator-selection policy online for this routing layer.
LA-BHH treats 2-opt, swap, relocate, and Or-opt moves as low-level arms, builds context from static landscape descriptors and online search-state features, and updates a LinUCB controller from improvement rewards during the same run.
Experimental results on 45 generated Euclidean TSP instances show that LA-BHH achieves the best mean final gap and convergence AUC, with 0.0223 and 0.0389 respectively.
It reduces final gap by 17.6\% over UCB-HH, 22.6\% over Random-HH, and 68.2\% over nearest-neighbor construction.
Ablation results further show that contextual credit assignment, 2-opt repair, and stagnation-aware state use are the main contributors.
\end{abstract}

\section{Introduction}

Learning-assisted algorithm design (LEAD) studies how optimization algorithms can adapt their own design choices through data, feedback, and search experience.
Instead of committing to one manually fixed update rule or operator schedule, a LEAD method can use instance descriptors, online performance traces, or learned policies to decide which algorithmic action should be taken next.
This viewpoint has become increasingly relevant across learning to optimize, neural combinatorial optimization, meta black-box optimization, data-driven evolutionary algorithms, and reinforcement-learning-assisted operator selection.
For many practical optimization tasks, however, the most useful form of learning is not necessarily a large offline model.
A compact online controller that learns during a single run can be easier to inspect, cheaper to reproduce, and more suitable when only a small number of related instances are available.

This paper focuses on UAV multi-site inspection route planning.
The setting appears in campus roof and facade inspection, industrial park patrol, power-line or pipeline inspection, agricultural sensing, and disaster-site reconnaissance.
An inspection sortie must visit a set of assigned sites and return to the depot within a limited flight window, making the visiting order the dominant decision when altitude control, collision avoidance, and service times are handled by lower-level planners.
This route-ordering view is compatible with recent UAV navigation work that separately addresses safe local motion and goal-aware planning~\cite{wei2026saga}.
Euclidean traveling-salesman instances provide a clean abstraction for this setting, while generated landscape families make it possible to test whether an algorithm remains stable across open-area, clustered, corridor-like, grid-like, and mixed-density deployments.
Routing-oriented studies have reached a similar practical lesson from adjacent settings: adaptive position updates for UAV path planning and geometric swarm search improve performance by shaping how candidate moves are generated and repaired~\cite{wei2024apupso,wei2026geossa}.
Related engineering-design studies also show that triangular walks, multi-strategy whale search, and geometric flight mechanisms can strengthen black-box search by changing the update mechanism itself~\cite{wei2025tswoa,gu2025gwoa,wei2026gwo}.
Here we move this design issue one level upward.
The low-level moves remain standard routing neighborhoods, while learning is used to decide which operator should be applied under the current landscape and search state.

Classical routing heuristics remain attractive because they are simple, fast, and interpretable.
Nearest-neighbor construction provides a strong starting point, 2-opt repairs crossing edges, relocate and Or-opt moves correct local ordering, and swap moves can exchange misplaced sites across clusters.
These operators have different strengths across different landscapes and different stages of search.
A fixed schedule may overuse an operator after it stops improving, while a purely random hyper-heuristic may waste evaluations on moves that are unlikely to help under the current state.
This motivates an adaptive selection mechanism that combines static information about the instance with dynamic information about search progress, recent improvement, acceptance rate, and stagnation.

Hyper-heuristics provide a natural framework for this goal because they search over heuristics instead of directly searching only over candidate solutions~\cite{burke2013hyper}.
In a selection hyper-heuristic, low-level operators remain ordinary optimization moves, while a high-level controller decides which operator to apply at each iteration.
This separation is well aligned with LEAD: the low-level search components preserve domain knowledge, and the high-level policy learns how to use them under changing conditions.
Contextual bandits are a particularly lightweight choice for such control, since they can trade off exploration and exploitation while conditioning decisions on feature vectors~\cite{li2010contextual}.

We propose LA-BHH, a landscape-aware bandit hyper-heuristic for adaptive operator selection in Euclidean UAV inspection routing.
LA-BHH uses static landscape descriptors and online search-state features as the context for a LinUCB controller over four low-level neighborhoods: 2-opt, swap, relocate, and Or-opt.
The method requires no offline training set, neural architecture, or external solver.
Its learning signal is the immediate tour-length improvement produced by the selected operator, which is converted into a reward and used to update the corresponding operator credit during the same run.
It also uses a simple stagnation-aware repair gate to bias selection toward 2-opt when the search stops improving.
This design keeps the algorithm close to practical routing heuristics while introducing an explicit learning component that can be evaluated through ablation.

The experimental study compares LA-BHH with nearest neighbor, stochastic 2-opt, simulated annealing, iterated local search, a genetic algorithm, Random-HH, and non-contextual UCB-HH.
The benchmark covers five generated landscape families and three instance sizes, with convergence curves, final gaps, route visualizations, landscape-wise heatmaps, operator-usage analysis, and component-removal ablations.
The results show that LA-BHH improves final solution quality and convergence behavior over the default baselines, while the ablations clarify which parts of the design matter most.
Together, these results position LA-BHH as a compact LEAD case study for hyper-heuristics, reinforcement-learning-assisted operator selection, and landscape-aware UAV inspection routing.

\section{Related Work}

\paragraph{Hyper-heuristics and adaptive operator selection.}
Hyper-heuristics raise the level of search control from solution moves to heuristic selection or generation~\cite{burke2013hyper,ross2005hyper,drake2020selection}.
Selection hyper-heuristics are especially relevant to routing because neighborhoods such as 2-opt, swap, relocation, and segment moves have complementary strengths across spatial layouts.
Adaptive large-neighborhood search and related routing heuristics also update operator utilities online~\cite{ropke2006adaptive,pisinger2007general}, but they are often engineered around problem-specific destroy-and-repair mechanisms.
LA-BHH keeps the operator portfolio deliberately simple and uses a contextual bandit to make the credit assignment mechanism explicit.

\paragraph{Learning-assisted algorithm design.}
Algorithm selection was formalized by Rice~\cite{rice1976algorithm} and later developed through empirical configuration methods such as racing, SMAC, and irace~\cite{birattari2002racing,hutter2011sequential,lopez2016irace}.
Surveys on meta-learning and automated algorithm selection emphasize the role of instance descriptors and algorithm-behavior traces~\cite{smith2009cross,kerschke2019automated}.
Exploratory landscape analysis provides compact descriptors for global structure, dispersion, and search difficulty~\cite{mersmann2011exploratory,mueller2015exploratory}.
Our method adopts this LEAD view but keeps learning online: static descriptors are combined with short-horizon search-state features during a single run.

\paragraph{Bandits and routing heuristics.}
Multi-armed bandits provide simple exploration--exploitation mechanisms with finite-time analysis~\cite{auer2002finite}, and contextual variants such as LinUCB use feature vectors when estimating action value~\cite{li2010contextual}.
Bandit-style upper-confidence control is also central to UCT in Monte-Carlo tree search~\cite{kocsis2006bandit}.
For routing, 2-opt and Lin--Kernighan remain core local-search mechanisms~\cite{croes1958method,lin1973effective,helsgaun2000effective}, and TSPLIB has long served as a standard benchmark collection~\cite{reinelt1991tsplib,applegate2006traveling}.
LA-BHH does not attempt to replace highly optimized TSP solvers; instead, it studies how a compact online controller can select among reusable neighborhoods under controlled landscape variation.

\section{Problem Setting}

We consider a Euclidean traveling-salesman problem (TSP).
Given coordinates $X=\{x_1,\ldots,x_n\}$, the objective is to find a permutation $\pi$ minimizing
\begin{equation}
    f(\pi)=\sum_{i=1}^{n} d(x_{\pi_i},x_{\pi_{i+1}}),
\end{equation}
where $\pi_{n+1}=\pi_1$ and $d(\cdot,\cdot)$ is Euclidean distance.
The target application is UAV multi-site inspection: an inspection sortie starts from a depot, visits all assigned sites, and returns to the depot under a limited flight window.
This abstraction covers campus roof and facade inspection, industrial park patrol, agricultural sensing-point collection, power-line or pipeline inspection, and disaster-site reconnaissance when low-level flight control and obstacle avoidance are secondary to visit ordering.

The benchmark generator covers five instance-space families: uniform, clustered, corridor, grid-jitter, and mixed-density.
These families correspond to common deployment layouts: uniform tasks in open areas, clustered task sites in farms or industrial zones, corridor-like routes along roads or pipelines, grid-jitter layouts on campuses, and mixed-density urban--suburban deployments.
They are not intended to replace TSPLIB~\cite{reinelt1991tsplib}; rather, they provide controlled landscape variation for studying adaptive operator selection under application-like routing structure.

\section{LA-BHH}

LA-BHH maintains one incumbent tour and a portfolio of four low-level operators: 2-opt, swap, relocate, and Or-opt-2.
At each iteration, the controller observes a feature vector and selects one operator.
The resulting candidate is evaluated by a greedy best-improvement acceptance rule and converted into a reward for the selected operator.
This reward closes the online learning loop: operators that produce larger improvements under similar contexts receive higher estimated value, while the confidence term keeps unexplored operators available.
When the search stagnates, LA-BHH uses the dynamic state to bias the controller toward 2-opt repair moves, which gives the online state features a direct exploitation role rather than using them only as noisy regression inputs.
Algorithm~\ref{alg:labhh} summarizes the resulting online selection loop.

\subsection{Landscape and State Features}

The static landscape vector contains normalized problem size, mean nearest-neighbor distance, nearest-neighbor dispersion, coordinate anisotropy, radial dispersion, and MST weight per node.
The dynamic vector contains budget progress, current and best tour length relative to the initial tour, recent improvement, recent acceptance rate, and stagnation.
The full context is
\begin{equation}
    z_t = [1, s_{\mathrm{static}}, s_{\mathrm{dynamic}}].
\end{equation}

\subsection{Bandit Controller}

For each operator $a$, LA-BHH keeps ridge-regression statistics $(A_a,b_a)$.
At iteration $t$, it selects
\begin{equation}
    a_t=\arg\max_a \hat{\theta}_a^\top z_t
    + \alpha \sqrt{z_t^\top A_a^{-1}z_t},
\end{equation}
where $\hat{\theta}_a=A_a^{-1}b_a$.
The reward is the clipped relative improvement produced by the candidate move:
\begin{equation}
    r_t=\mathrm{clip}\left(100\frac{f(\pi_t)-f(\pi'_t)}{f(\pi_0)}, -1, 1\right).
\end{equation}
The selected arm is updated with $A_a \leftarrow A_a + z_tz_t^\top$ and $b_a \leftarrow b_a+r_tz_t$.
This update is the learning step in LA-BHH: the controller changes its future operator choices from the rewards observed during the current optimization run.

\subsection{Low-Level Operators and Acceptance}

The operator portfolio contains four permutation-preserving moves.
The 2-opt operator reverses a tour segment and directly targets edge crossings, swap exchanges two inspection sites, relocate moves one site to another position, and Or-opt-2 relocates a two-site block.
For each selected operator, LA-BHH samples three candidate moves and keeps the best candidate under the current tour length.
This small candidate pool reduces the variance of single random moves while preserving the same operator-selection interface for all methods.

The initial incumbent is the best tour found by multi-start nearest neighbor over all possible starts.
This deliberately makes the task harder for the learned controller: improvements must come from adaptive local search rather than from weak construction.
Candidate acceptance is greedy by default.
Annealing-style acceptance is included as an ablation, but it degraded final performance in these Euclidean inspection instances because worsening moves often disrupted already strong nearest-neighbor structure.

\subsection{Stagnation-Aware State Use and Complexity}

The online state is used in two ways.
First, it enters the LinUCB context, allowing the controller to condition operator value estimates on search progress, improvement rate, acceptance rate, and stagnation.
Second, when the search has not improved for a short window, the state triggers a repair gate that increases the probability of choosing 2-opt.
This conservative design does not introduce a new move; it biases selection toward an operator known to be effective for Euclidean routing repair.

The per-iteration cost is dominated by evaluating a constant number of candidate tours.
With direct tour-length evaluation, this is $O(n)$ per candidate and $O(Tn)$ for $T$ iterations.
The LinUCB update has constant dimension in this study, so its cost is negligible relative to candidate evaluation for the problem sizes considered.
The resulting computational cost is suitable for medium-size routing benchmarks without specialized solvers or GPU acceleration.

\begin{algorithm}[t]
\caption{LA-BHH}
\label{alg:labhh}
\begin{algorithmic}[1]
\STATE Build an initial tour $\pi_0$ by nearest neighbor
\STATE Compute static landscape features
\FOR{$t=1,\ldots,T$}
    \STATE Build context $z_t$ from static and dynamic features
    \STATE Select operator $a_t$ by LinUCB
    \STATE If stagnation is detected, bias $a_t$ toward 2-opt repair
    \STATE Apply $a_t$ to obtain candidate $\pi'_t$
    \STATE Evaluate $f(\pi'_t)$ and compute reward $r_t$
    \STATE Accept the candidate if it improves the incumbent
    \STATE Update the selected LinUCB arm with $(z_t,r_t)$
\ENDFOR
\STATE \textbf{return} the best tour found
\end{algorithmic}
\end{algorithm}

\section{Experiments}

\subsection{Experimental Setup}

\begin{figure*}[t]
    \centering
    \includegraphics[width=\textwidth]{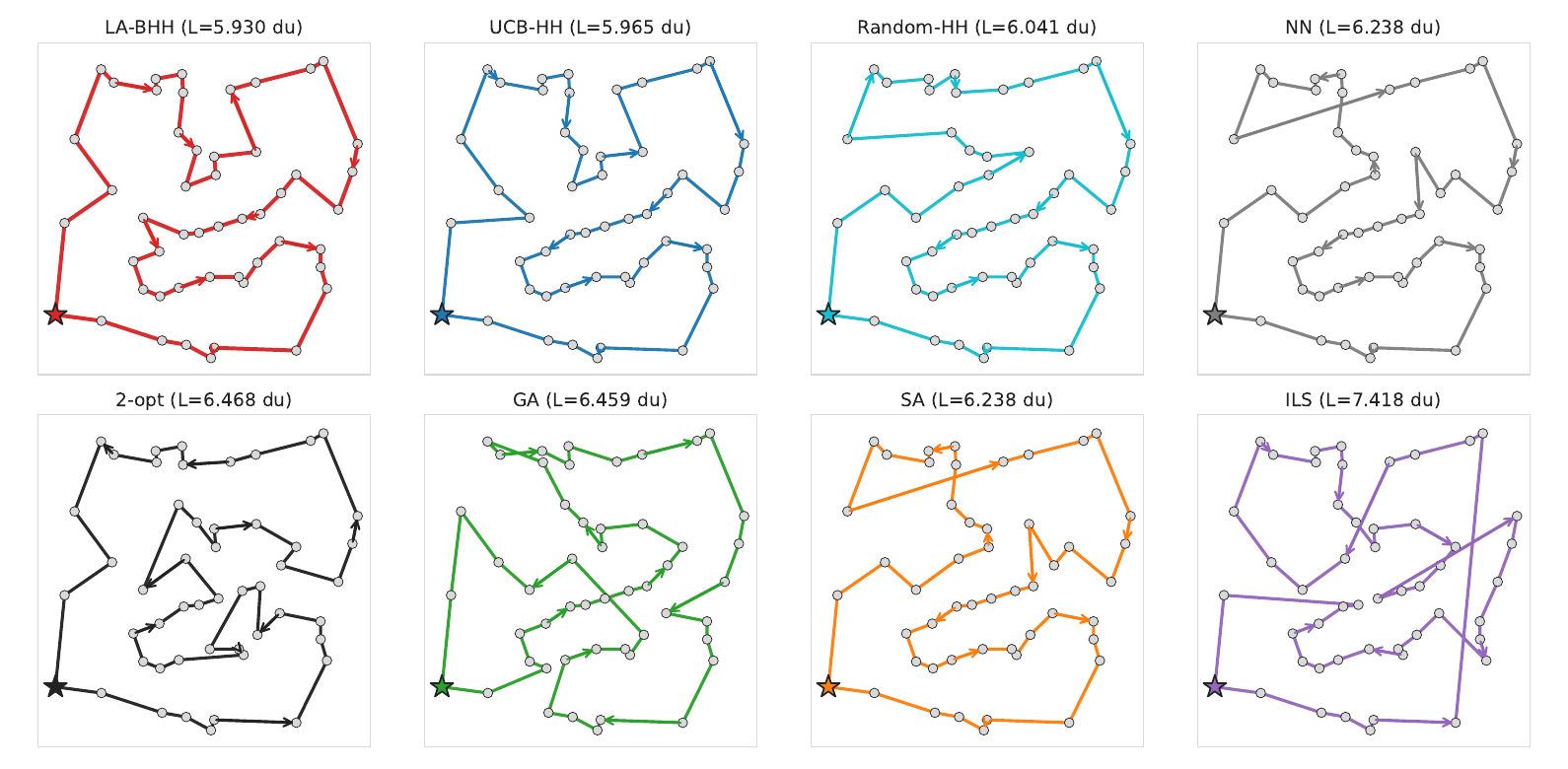}
    \caption{UAV multi-site inspection routes produced by all compared algorithms on the same generated instance. Stars mark the depot, gray circles mark inspection sites, and panel titles report route length $L$ in normalized distance units (du).}
    \label{fig:inspection-route}
\end{figure*}

We compare LA-BHH against nearest neighbor~\cite{rosenkrantz1977analysis}, stochastic 2-opt~\cite{croes1958method}, simulated annealing~\cite{kirkpatrick1983optimization}, iterated local search~\cite{lourenco2003iterated}, a genetic algorithm~\cite{muhlenbein1992parallel}, Random-HH~\cite{burke2013hyper}, and non-contextual UCB-HH~\cite{auer2002finite}.
The set separates construction quality, local repair, trajectory-based metaheuristic search, population-based search, random operator selection, and bandit credit assignment without contextual conditioning.
The evaluation uses instance sizes $n\in\{50,100,200\}$, five landscape families, three instances per family-size pair, and three random seeds.
This gives 45 generated instances for comparing LA-BHH with seven default baselines and nine component-removal variants.
All experiments were conducted on a Linux workstation with an Intel Xeon E5-2698 v4 CPU, 20 physical cores and 40 hardware threads at 2.20GHz, and 251GB RAM.
The workstation also contains four NVIDIA Tesla V100 32GB GPUs, but all reported routing algorithms are CPU-only and do not use GPU acceleration.
The software environment used Python 3.13.12 and NumPy 2.4.4 for the optimization runs.
The main metrics are final relative gap to the best method observed on the same instance, convergence AUC, and runtime.
Figure~\ref{fig:inspection-route} visualizes the target UAV inspection scenario and representative routes from compared methods.

The ablation suite removes one design component at a time: all context, dynamic search-state features, static landscape features, annealing-style acceptance, and learned credit assignment.
It also drops each low-level operator independently.
This directly tests whether performance comes from contextual learning, online state adaptation, controlled diversification, or a single dominant operator.

For each instance, the reference value in the gap metric is the best tour length observed among all compared methods, variants, and seeds for that instance.
This avoids claiming optimality while still enabling fair within-instance comparison.
Convergence AUC is computed from logged best-so-far gaps over the budget; lower AUC means the method finds good tours earlier.
Runtime is reported only as an auxiliary diagnostic, since all methods use the same simple data structures rather than optimized solver kernels.

\subsection{Overall Performance}

\begin{table}[t]
\centering
\caption{Mean performance over 45 generated instances and three random seeds. Gap is measured relative to the best tour observed for the same instance, AUC summarizes best-so-far convergence, and runtime is CPU time in seconds. Lower values are better.}
\label{tab:main}
\small
\begin{tabular}{lccc}
\toprule
Algorithm & Gap & AUC & Runtime (s)\\
\midrule
LA-BHH & $\boldsymbol{0.0223}$ & $\boldsymbol{0.0389}$ & 1.314\\
UCB-HH & 0.0270 & 0.0437 & 0.973\\
Random-HH & 0.0287 & 0.0444 & 0.702\\
NN & 0.0700 & 0.0700 & 0.153\\
2-opt & 0.0961 & 0.1213 & $\boldsymbol{0.120}$\\
GA & 0.1149 & 0.1384 & 0.491\\
SA & 0.1522 & 0.1611 & 0.185\\
ILS & 0.1574 & 0.1582 & 0.121\\
\bottomrule
\end{tabular}
\end{table}

\begin{figure}[t]
    \centering
    \includegraphics[width=\linewidth]{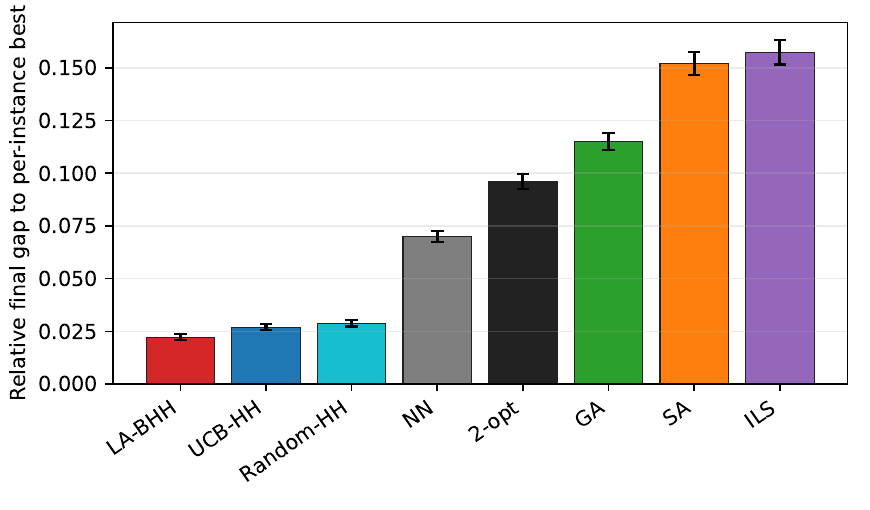}
    \caption{Mean relative final gap of the compared algorithms. Error bars denote the standard error over all runs.}
    \label{fig:main-gap}
\end{figure}

Table~\ref{tab:main} shows that adaptive hyper-heuristics dominate the classical baselines on this benchmark.
LA-BHH obtains the best final gap and convergence AUC among the compared methods, improving the final gap by 17.6\% over UCB-HH, 22.6\% over Random-HH, and 68.2\% over nearest-neighbor construction.
This indicates that contextual credit assignment and stagnation-aware repair provide useful guidance beyond random operator mixing and non-contextual bandit selection.
Figure~\ref{fig:main-gap} visualizes the aggregate final-gap comparison.
The result is not driven by one weak baseline: LA-BHH improves over construction, local-search, evolutionary, random hyper-heuristic, and non-contextual bandit baselines under the same instance set.

\begin{figure}[t]
    \centering
    \includegraphics[width=\linewidth]{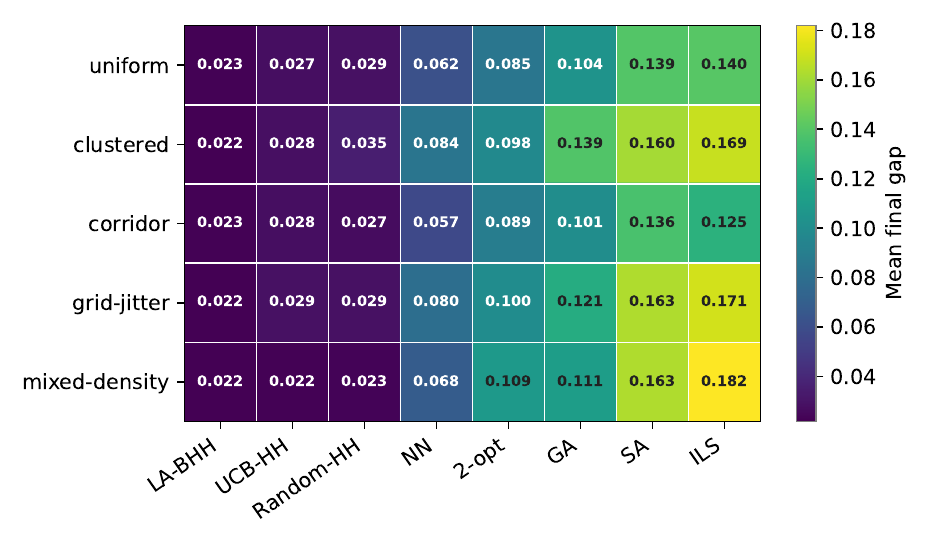}
    \caption{Mean relative final gap by landscape family and algorithm. Lower values indicate better final tours within each generated instance family.}
    \label{fig:heatmap}
\end{figure}

Figure~\ref{fig:heatmap} breaks the final-gap comparison down by landscape family.
LA-BHH is consistently among the strongest methods across uniform, clustered, corridor, grid-jitter, and mixed-density instances.
The mixed-density family is the closest case, where UCB-HH reaches a similar mean final gap, but the contextual controller retains the best overall average.
The heatmap therefore supports the central claim that the proposed controller improves robustness across different routing landscapes rather than only improving one instance family.

\subsection{Ablation Study}

\begin{figure}[t]
    \centering
    \includegraphics[width=\linewidth]{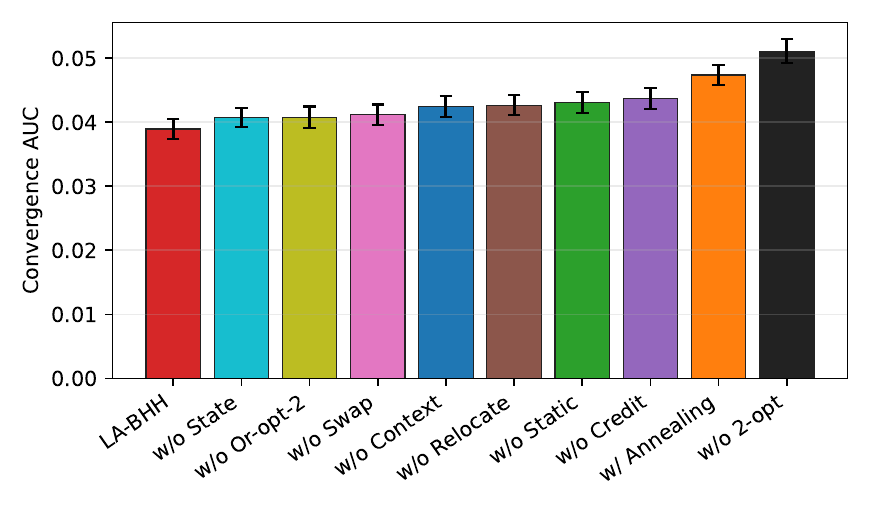}
    \caption{Component ablation of LA-BHH measured by convergence AUC. Labels with w/o remove one component or operator, while w/ Annealing replaces greedy acceptance with annealing-style acceptance. Lower values indicate faster convergence to high-quality tours.}
    \label{fig:ablation}
\end{figure}

The ablation results in Figure~\ref{fig:ablation} clarify which components are responsible for the observed gains.
Removing 2-opt causes the largest degradation, confirming that the controller depends on a strong Euclidean repair operator rather than gaining from portfolio diversity alone.
Adding annealing-style acceptance degrades performance on these instances, so the default LA-BHH uses greedy acceptance.
Removing context, static landscape features, or learned credit assignment also worsens AUC, which indicates that the bandit controller benefits from both landscape information and online reward learning.
Removing the online state has a smaller effect on final gap but worsens convergence behavior, making the state signal most useful for early progress through the stagnation-aware repair gate.

\subsection{Convergence, Scaling, and Operator Use}

\begin{figure}[t]
    \centering
    \includegraphics[width=\linewidth]{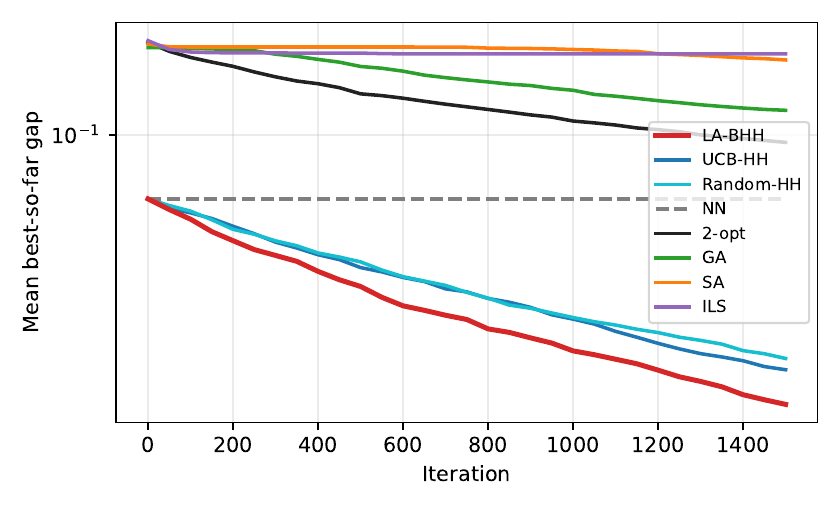}
    \caption{Mean best-so-far relative gap across the shared evaluation budget. Nearest-neighbor construction is shown as a constant reference because it does not perform iterative improvement. Lower curves indicate faster convergence.}
    \label{fig:convergence}
\end{figure}

Figure~\ref{fig:convergence} shows that LA-BHH reaches lower gaps earlier under the shared evaluation budget.
The gap between LA-BHH and Random-HH is visible throughout most of the run, showing that the learned controller is not only improving the final incumbent but also changing the search trajectory.
UCB-HH follows a similar trend but remains less stable because it does not condition arm values on landscape and search-state features.

\begin{figure}[t]
    \centering
    \includegraphics[width=\linewidth]{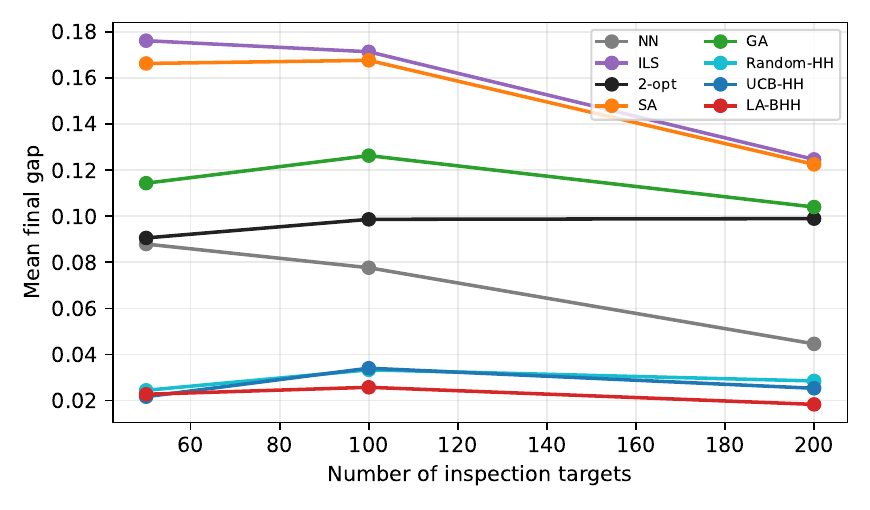}
    \caption{Mean relative final gap as the number of inspection sites increases from 50 to 200. Lower values indicate better scalability in final tour quality.}
    \label{fig:size-scaling}
\end{figure}

Figure~\ref{fig:size-scaling} reports the same final-gap metric as the number of inspection sites increases.
LA-BHH remains competitive from 50 to 200 sites, while classical local-search baselines become less reliable as the tour grows.
This behavior is consistent with the design goal of learning when to apply repair and relocation operators instead of relying on a fixed local-search pattern.

\begin{figure}[t]
    \centering
    \includegraphics[width=\linewidth]{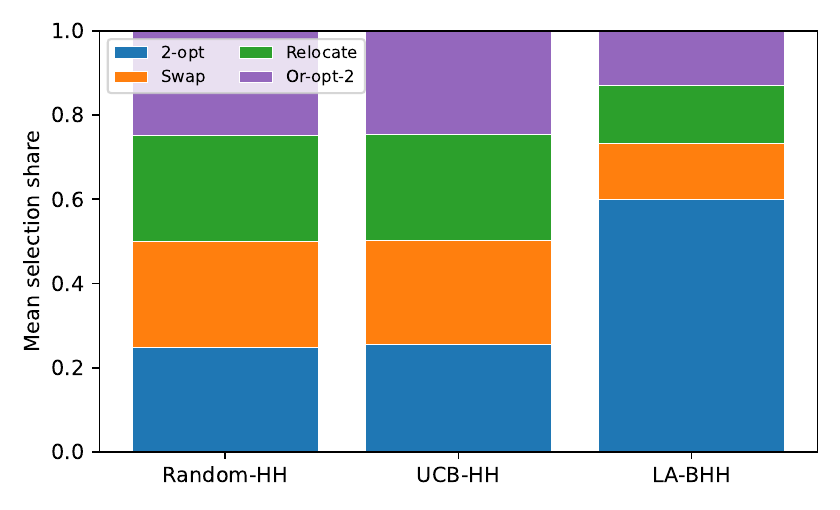}
    \caption{Mean low-level operator selection share for hyper-heuristic methods. Segment colors denote low-level operators, and shares sum to one within each method.}
    \label{fig:operator-usage}
\end{figure}

Figure~\ref{fig:operator-usage} compares how the hyper-heuristic methods allocate the same low-level operator portfolio.
Random-HH spreads decisions nearly uniformly by design, while UCB-HH and LA-BHH assign more probability mass to operators that have accumulated stronger rewards.
LA-BHH still keeps nonzero use of swap, relocate, and Or-opt-2, which is important because these moves can repair local ordering errors that 2-opt does not directly address.
The usage pattern matches the ablation evidence: 2-opt is the dominant repair move, but the full portfolio contributes to stable performance.

\section{Discussion}

The results support a practical interpretation of learning-assisted algorithm design for routing.
LA-BHH is not a replacement for a state-of-the-art TSP solver such as LKH; rather, it is a compact controller for deciding which reusable search operator should be applied under a fixed budget.
This distinction matters in LEAD settings where the goal is to learn reusable algorithm-design decisions rather than to hand-engineer one specialized solver.
The strongest gains occur when the controller combines a good construction heuristic, explicit credit assignment, and a repair bias triggered by stagnation.

The ablations also show that not every learning signal is automatically useful.
Dynamic state features improve convergence AUC, but their effect on the final best tour can be small because the search budget is short and the multi-start nearest-neighbor initialization is already strong.
This suggests that future work should explore more stable state summaries, such as edge-frequency entropy, local crossing counts, or recent operator-specific success rates.
Another extension is to use the same hyper-heuristic controller for hyperparameter and operator configuration in deep learning prediction pipelines, following recent optimization-assisted models for academic-potential prediction and stock-price forecasting~\cite{wei2026nawoa,li2026askssa}.
Another limitation is that the generated instances are controlled rather than field-collected.
They are appropriate for isolating landscape effects, but deployment studies on real UAV inspection maps with no-fly zones, altitude constraints, service times, and battery constraints would be needed before operational use.
For practical UAV inspection systems, the present formulation should be viewed as the route-ordering layer after candidate inspection sites have been extracted from a map.
Obstacle avoidance, localization uncertainty, charging constraints, and service-time windows can be handled by a lower-level motion planner, such as heuristic RRT-style urban planning~\cite{ahrrt}, or by extending the tour cost with additional penalties.
This separation is useful because LA-BHH does not require the low-level operators to know every deployment detail; the high-level controller only needs reliable feedback on whether an operator improves the current routing objective.
The same structure can therefore be transferred to field maps by replacing Euclidean distances with shortest-path distances on a road, aisle, or traversability graph.
Coverage-oriented inspection can be handled in the same way by replacing route length with a weighted route--coverage objective, following coverage optimization studies for wireless sensor networks~\cite{wei2025lnwoa}.

The current controller also leaves room for stronger learning mechanisms without changing the hyper-heuristic interface.
Cross-instance memory could initialize the bandit parameters from previous inspection maps, while online updates would still adapt to the current run.
Richer state features, such as edge-frequency entropy, local crossing counts, or operator-specific recent success rates, may help distinguish stagnation caused by local optima from stagnation caused by poor operator choice.
These directions fit the LEAD objective of learning reusable algorithm-design decisions while keeping the resulting optimizer transparent enough for routing applications.

\section{Conclusion}

This paper presented LA-BHH, a landscape-aware bandit hyper-heuristic for adaptive operator selection in Euclidean UAV inspection routing.
The method uses static landscape descriptors and online search-state features to guide a contextual bandit over a small portfolio of routing neighborhoods.
Experiments across five generated landscape families show that LA-BHH improves final solution quality and convergence behavior over classical heuristics, random operator selection, and a non-contextual UCB hyper-heuristic.
The ablation study further indicates that the gains come mainly from the interaction between 2-opt repair, contextual credit assignment, and stagnation-aware state use.
These results support lightweight online learning as a practical design pattern for adaptive UAV inspection routing, while leaving richer state representations and real inspection maps as the next steps.

\clearpage
\bibliographystyle{named}
\bibliography{references}

\end{document}